\documentclass[preprint2,twocolappendix,trackchanges]{aastex6}

\usepackage{booktabs}
\usepackage{color}
\usepackage{soul}
\usepackage{ulem}
\usepackage{bm}
\usepackage{amsmath}

\def\pj#1{{\color{red} #1}}

\begin{document}


\title{Extended Calculations of Spectroscopic Data:  Energy Levels, Lifetimes and Transition rates for O-like ions from  C\lowercase{r}~XVII to Z\lowercase{n}~XXIII}

\author{K. Wang\altaffilmark{1,2}, P. J\"{o}nsson\altaffilmark{2}, J. Ekman\altaffilmark{2}, G. Gaigalas\altaffilmark{3}, M. R. Godefroid\altaffilmark{4}, \\R. Si\altaffilmark{5}, Z.B. Chen\altaffilmark{6}, S. Li\altaffilmark{7}, C.Y. Chen\altaffilmark{5}, and J. Yan\altaffilmark{7,8,9}}


\affil{$^1$Hebei Key Lab of Optic-electronic Information and Materials, The College of Physics Science and Technology, Hebei University, Baoding 071002, China\\
	$^2$Group for Materials Science and Applied Mathematics, Malm\"o University, SE-20506, Malm\"o, Sweden; {\color{blue} per.jonsson@mah.se}\\
	$^3$Institute of Theoretical Physics and Astronomy, Vilnius University, Saul\.etekio av. 3, LT-10222, Vilnius, Lithuania\\
	$^4$Chimie Quantique et Photophysique, CP160/09, Universit\'{e} libre de Bruxelles, Av. F.D. Roosevelt 50, 1050 Brussels, Belgium\\
    $^5$Shanghai EBIT Lab, Institute of Modern Physics, Department of Nuclear Science and Technology, Fudan University, Shanghai 200433, China; {\color{blue} chychen@fudan.edu.cn}\\
	$^6$College of Science,  National University of Defense Technology, Changsha 410073, China {\color{blue} chenzb008@qq.com}\\
	$^7$Institute of Applied Physics and Computational Mathematics, Beijing 100088, China;  {\color{blue} lishuangwuli@126.com} \\
	$^8$Center for Applied Physics and Technology, Peking University, Beijing 100871, China\\
    $^9$Collaborative Innovation Center of IFSA (CICIFSA), Shanghai Jiao Tong University, Shanghai 200240, China}






\begin{abstract}
Employing two state-of-the-art methods, multiconfiguration Dirac--Hartree--Fock and second-order many-body perturbation theory, the excitation energies and lifetimes for the lowest 200 states of the $2s^2 2p^4$, $2s 2p^5$, $2p^6$, $2s^2 2p^3 3s$, $2s^2 2p^3 3p$, $2s^2 2p^3 3d$, $2s 2p^4 3s$, $2s 2p^4 3p$, and $2s 2p^4 3d$ configurations, and multipole (electric dipole (E1), magnetic dipole (M1), and electric quadrupole (E2)) transition rates, line strengths, and oscillator strengths among these states are calculated for each O-like ion from \ion{Cr}{17} to \ion{Zn}{23}. Our two data sets are compared with the NIST and CHIANTI compiled values, and previous calculations. The data are accurate enough for identification and deblending of new emission lines from the sun and other astrophysical sources. The amount of data of high accuracy is significantly increased for the $n = 3$ states of several O-like ions of astrophysics interest, where experimental data are very scarce.  
\end{abstract}



\section{INTRODUCTION}\label{sect:in}
There is a wealth of observations from missions such as Chandra, XMM-Newton and Hinode.
These observations require theoretical studies to supply more extensive atomic data of high accuracy. Considering this, we have recently provided accurate atomic data for L-shell ions, including the beryllium, boron, carbon, nitrogen, fluorine, and neon isoelectronic sequences~\citep{Jonsson.2013.V559.p100,Ekman.2014.V564.p24,Wang.2014.V215.p26,Wang.2015.V218.p16,Wang.2016.V223.p3,Wang.2016.V226.p14,Radziute.2015.V582.p61,Si.2016.V227.p16}. In this paper, systematic calculations for oxygen-like ions from \ion{Cr}{17} to \ion{Zn}{23} are reported. 

Spectra of O-like ions, including the ions of the iron period, have been observed in various kinds of astronomical objects  such as the sun, distant stars, and the Milky Way~\citep{Fawcett.1987.V225.p1013,Feldman.1991.V75.p925,Landi.1997.V324.p1027,Feldman.1998.V503.p467,Feldman.2000.V544.p508,Behar.2001.V548.p966,Dere.2001.V134.p331,Mewe.2001.V368.p888,Kaastra.2002.V386.p427,Curdt.2004.V427.p1045,Landi.2005.V160.p286,Landi.2006.V166.p421,Raassen.2013.V550.p55}, as well as in laboratory plasmas~\citep{Brown.2002.V140.p589,Fournier.2003.V36.p3787,May.2005.V158.p230,Chen.2007.V168.p319,Traebert.2014.V211.p14,Yang.2016.V152.p135}. By analyzing lines of the spectra, properties of the plasmas, such as elemental abundances, electron density and temperature, can be determined. The $n=3 \rightarrow 2$ emission lines of \ion{Fe}{19} from the accumulated Chandra High Energy Transmission Grating (HETG) observations of Capella, for instance, were used for temperature and density diagnostics of the corona of Capella~\citep{Canizares.2000.V539.p41,Kotochigova.2010.V186.p85}. Ni lines have been identified in the solar X-ray spectra~\citep{Phillips.1982.V256.p774}, and in the central region of the Perseus cluster~\citep{Tamura.2009.V705.p62}, and offer additional information on the emission measure distribution and density. 

Many calculations have been performed for O-like ions. Most of them 
targeted atomic data for low-lying states of the $(1s^2) 2s^2 2p^4$, $2s 2p^5$, and $2p^6$ configurations (the $n=2$ complex)~\citep{Baluja.1988.V21.p15,Baluja.1988.V21.p1455,Galavis.1997.V123.p159,Vilkas.1999.V60.p2808,Zhang.2002.V82.p357,Gu.2005.V89.p267,Hu.2011.V9.p1228,Rynkun.2013.V557.p136,Fontes.2015.V101.p143}. 

Because of their wide applications for analyzing new observations of astrophysical sources, as well as for modeling and diagnosing a variety of plasmas, energy and transition data for higher-lying states of the $n \geq 3$ complexes are also eagerly needed~\citep{Phillips.1982.V256.p774,Acton.1985.V291.p865,Landi.2005.V160.p286,Kotochigova.2007.V76.p52513,Kotochigova.2010.V186.p85,Raassen.2013.V550.p55}. Among the studies of the $n \geq 3$ states of \ion{Fe}{19} we mention the calculations of~\citet{Jonauskas.2004.V424.p363} using the multiconfiguration Dirac-Hartree-Fock (MCDHF) method, the calculations of~\citet{Landi.2006.V640.p1171} using the standard relativistic configuration interaction (RCI) method, the AUTOSTRUCTURE calculations of~\citet{Butler.2008.V489.p1369}, and the relativistic Breit--Pauli calculations of~\citet{Nahar.2011.V97.p403}. MCDHF calculations for \ion{Ni}{21} were performed by~\citet{Fan.2013.V67.p255} for the lowest 86 states of the $n \leq 3$ configurations using the GRASP89 code~\citep{Dyall.1989.V55.p425}, and by~\citet{Aggarwal.2014.V92.p1285} for the lowest 200 states of the $n \leq 3$ configurations using the GRASP0 code~\citep{Grant.1980.V21.p207}. Using a configuration interaction Dirac--Fock and Dirac--Fock--Sturm method combined with second-order Brillouin--Wigner perturbation theory, ~\citet{Kotochigova.2007.V76.p52513,Kotochigova.2010.V186.p85} computed the wavelengths and oscillator strengths for the $2s^2 2p^3 3s,3d \rightarrow 2s^2 2p^4$ and $2s 2p^4 3p \rightarrow 2s^2 2p^4$ emission lines in the wavelength range from  12 \AA~ to \mbox{16 \AA~} with high accuracy. Excitation energies of the $2l^6$ and $2l^5 3l'$ states for \ion{Fe}{19} and \ion{Ni}{21}, and wavelengths of 
the $2l^5 nl'$, $n \rightarrow 2$ (with $3 \leq n \leq 7$) transitions for the same ions were reported by~\citet{Gu.2005.V156.p105,Gu.2007.V169.p154} using a  combined RCI and many-body perturbation theory (MBPT) method. 

Energy and transition data of the $n=2$ states provided by 
~\citet{Vilkas.1999.V60.p2808,Gu.2005.V89.p267,Rynkun.2013.V557.p136}  are of high accuracy and can be used to identify observed spectral lines among the $n=2$ states. The data sets of the $n \geq 3$ states reported by~\citet{Gu.2005.V156.p105,Gu.2007.V169.p154,Kotochigova.2007.V76.p52513,Kotochigova.2010.V186.p85} are also of high accuracy. 
However, \citet{Gu.2005.V156.p105,Gu.2007.V169.p154} provided energy data for O-like Fe and Ni but not transition rates, and ~\citet{Kotochigova.2007.V76.p52513,Kotochigova.2010.V186.p85} only reported transition wavelengths and rates in the range from  12 \AA~ to 16 \AA~in \ion{Fe}{19}. In comparison, the calculations 
by~\cite{Landi.2006.V640.p1171}, ~\citet{Butler.2008.V489.p1369}, and ~\citet{Nahar.2011.V97.p403} are quite inaccurate because of limited configuration interaction effects, though they provide complete sets of data including transition rates. For example, the energy values of ~\cite{Landi.2006.V640.p1171}, ~\citet{Butler.2008.V489.p1369}, and ~\citet{Nahar.2011.V97.p403} for \ion{Fe}{19} depart from the compiled values in the Atomic Spectra Database (ASD) of the National Institute of Standards and Technology (NIST)~\citep[http://physics.nist.gov/asd]{Kramida.2015.V.p}, as well as the present calculated values, by up to 1.2~\%, 1.1~\%, and 2.9~\%, respectively.
These gaps are too large for identification and deblending of emission lines in collisionally ionized plasmas such as the solar corona.

The present work provides consistent data sets of energy structures and transition characteristics with high accuracy for  O-like ions in the range of nuclear charges $ 24 \le Z \le 30$.
Employing two state-of-the-art methods, the MCDHF and RCI method implemented in the latest version of the GRASP2K code~\citep{Jonsson.2013.V184.p2197}, and a combined RCI and MBPT approach in the FAC package~\citep{Gu.2008.V86.p675}, the excitation energies and lifetimes of the lowest 200 states for the $2s^2 2p^4$, $2s 2p^5$, $2p^6$, $2s^2 2p^3 3s$, $2s^2 2p^3 3p$, $2s^2 2p^3 3d$, $2s 2p^4 3s$, $2s 2p^4 3p$, and $2s 2p^4 3d$ configurations, and multipole transition rates (electric dipole (E1), magnetic dipole (M1), and  electric quadrupole (E2)) among these states are calculated for each ion from \ion{Cr}{17} to \ion{Zn}{23}.  The data sets obtained by the two theoretical methods are in excellent agreement. Compared with previous studies of O-like ions, our calculations result in a significant extension of accurate energy and transition data for higher-lying states of the $n = 3$ configurations, which will greatly improve the assessment of blending for diagnostic lines of interest, and aid the analysis of new spectra from astrophysical sources.
\section{Theory and Calculations}
\subsection{MCDHF}\label{Sec:MCDHF}
In relativistic theory an atomic state is described by a wave function, which is a solution to the wave equation based on the Dirac-Coulomb Hamiltonian. In the MCDHF method, the wave function $\Psi (\gamma PJM)$ for a state labeled $\gamma PJM$ with  $\gamma$ being the orbital occupancy and angular coupling tree quantum numbers, $P$ the parity, $J$ the total angular momentum quantum number, and $M$ the total magnetic quantum number, is expanded over configuration state functions $|\gamma_rPJM\rangle$ (CSFs)
\begin{equation}
	|\Psi(\gamma PJM)\rangle=\sum^{\rm NCSFs}_{r=1}c_r|\gamma_r P J 
	M\rangle\pj{.}
\end{equation}
The CSFs are antisymmetrized and symmetry-adapted many electron functions built from products of one-electron Dirac orbitals~\citep{Grant.2007.V.p}.  Based on the extended optimal level (EOL) scheme, the radial parts of the Dirac orbitals and the expansion coefficients of the targeted states are optimized to self-consistency by solving the MCDHF equations, which are derived using the variational approach. The Breit interaction and leading QED effects (vacuum polarization and self-energy) are included in subsequent RCI calculations, where the best expansion coefficients are determined for the frozen one-electron orbital set. Transition parameters such as transition rates $A$, line strengths $S$ or weighted oscillator strengths $gf$ between two states 
$\gamma PJM$ and $\gamma' P'J'M'$ are expressed in terms of the submatrix element of the transition operator
\begin{equation}
\langle \Psi(\gamma PJ) \| {\bf T} \| \Psi(\gamma' P'J') \rangle = \sum_{r,s} c_r c'_s \langle \gamma_r PJ \| {\bf T} \| \gamma'_s P'J' \rangle,
\end{equation}
where ${\bf T}$ is the transition operator~\citep{Cowan.1981.V.p}. The evaluation of the matrix elements follows the prescription given in \cite{Olsen.1995.V52.p4499}.

The MCDHF and RCI calculations for the O-like ions were based on a multireference single and double (MR-SD) process for 
generating CSF expansions and a systematic procedure for monitoring convergence of computed excitation energies and transition parameters~\citep{Fischer.2016.V49.p182004,Jonsson.2013.V559.p100,Ekman.2014.V564.p24}. The MR for the even states consisted of the $2s^22p^4$, $2p^6$, $2s^22p^33p$, $2s2p^43s$,
$2s2p^43d$, $2p^53p$, $2s^22p^34p$, $2s^22p^34f$, $2s2p^44s$, $2s2p^44d$ configurations. The MR for the odd states consisted of the
$2s2p^5$, $2s^22p^33s$, $2s^22p^33d$, $2s2p^43p$, $2p^53s$, $2p^53d$, $2s^22p^34s$, $2s^22p^34d$, $2s2p^44p$, $2s2p^44f$ configurations.
The CSFs were obtained by allowing SD substitutions from  the subshells of the configurations in the MR to an active set of orbitals that was extended to orbitals with quantum  numbers up to $n=8$ and $l = 6$. The substitutions were limited so that at most one substitution was allowed from the $1s^2$ core. For the even states there were 
6~787~000 CSFs distributed over the different $J$ symmetries whereas for the odd states there were 7~130~000 CSFs. The calculations where done by parity meaning that all the even parity states where determined together in one set of calculations and all odd parity states where determined together in another set of calculations. All calculations were performed with the GRASP2K code~\citep{Jonsson.2007.V177.p597,Jonsson.2013.V184.p2197}.


\subsection{MBPT}
A detailed description of the combined RCI and MBPT method can be found 
in~\citet{Lindgren.1974.V7.p2441,Safronova.1996.V53.p4036,Vilkas.1999.V60.p2808}.  ~\citet{Gu.2008.V86.p675}  implemented this method in the FAC package, which  has  successfully been used to calculate atomic data of high accuracy~\citep{Gu.2005.V156.p105,Gu.2007.V169.p154,Wang.2014.V215.p26,Wang.2015.V218.p16,Wang.2016.V223.p3,Wang.2016.V226.p14,Si.2016.V227.p16}.  In this method, the Hilbert space of the system is divided into two subspaces, including a model space $M$ and an orthogonal space $N$. 
By means of solving the eigenvalue problem of a non--Hermitian effective Hamiltonian in the space $M$, we can get the true eigenvalues of the Dirac--Coulomb--Breit Hamiltonian.  
The configuration interaction effects in the $M$ space is exactly considered, and the interaction of the spaces $M$ and $N$ is accounted for with the many-body perturbation theory up to the second order. 
In our calculations, we include all states of the  $2s^2 2p^4$, $2s 2p^5$, $2p^6$, $2s^2 2p^3 3s$, $2s^2 2p^3 3p$, $2s^2 2p^3 3d$, $2s 2p^4 3s$, $2s 2p^4 3p$, and $2s 2p^4 3d$ configurations in the model space \emph{M},
Through the SD virtual excitations of the states spanning the \emph{M} space, all states are contained in the space
\emph{N}. The maximum $n$ values for the single/double excitations are 200/65, respectively, while the maximum $l$ value is~20. Just as for the MCDHF and RCI calculations, QED corrections are also included.

\section{Evaluation of data}\label{sect:com}

\subsection{Energy Levels}\label{sect:en}
The computed excitation energies for all the 200 states of the $2s^2 2p^4$, $2s 2p^5$, $2p^6$, $2s^2 2p^3 3s$, $2s^2 2p^3 3p$, $2s^2 2p^3 3d$, $2s 2p^4 3s$, $2s 2p^4 3p$, and $2s 2p^4 3d$ configurations with $Z=24-30$ from our MCDHF/RCI and MBPT calculations are listed in Table~\ref{tab.lev}. In relativistic calculations the wave functions for the states are given as expansions over \emph{jj}-coupled CSFs. To provide the $LSJ$ labeling system used in databases such as the NIST ASD ~\citep{Kramida.2015.V.p} and CHIANTI~\citep{DelZanna.2015.V582.p56,Dere.1997.V125.p149}, the wave functions are
transformed  from a \emph{jj}-coupled CSF basis into a
$LSJ$-coupled CSF basis using the methods developed by~\citet{Gaigalas.2004.V157.p239}. For each state numbered by a key $(\#)$, the configuration and $LSJ$ designation,  the compiled values from the  NIST ASD when available, the radiative lifetime estimated from the theoretical transition rates (see section~3.2) and the eigenvector composition (largest expansion coefficients) are also included in Table~\ref{tab.lev} for the seven ions considered.

Since in astrophysics iron is of most concern, energies for all the 200 states from our MCDHF/RCI and MBPT calculations in \ion{Fe}{19} are compared with the compiled data from the NIST and CHIANTI databases in Table~\ref{tab.lev.Fe}. The previous theoretical values involving both the $n=2$ and $n=3$ states provided by~\citet{Landi.2005.V160.p286},~\citet{Butler.2008.V489.p1369} and~\citet{Nahar.2011.V97.p403} are also included. Due to the use of the same method and code, the MBPT energy data reported by~\cite{Gu.2005.V156.p105,Gu.2007.V169.p154} have a similar
accuracy as our MBPT values, and they are therefore not listed in Table~\ref{tab.lev.Fe}. 
Our MCDHF/RCI and MBPT excitation energies for \ion{Fe}{19} are in very good agreement.
Defining $\Delta E_i \equiv (E^i_{\rm MBPT} -E^i_{\rm MCDHF/RCI})$ for each of the $N$ energy levels that can be compared ($i=1,\ldots, N$), the average absolute difference between the two sets calculated from
\begin{equation}  \overline {\Delta E} 
= \frac{\sum \limits_{i=1}^{N} \Delta E_i}{N}  \; ,
\end{equation}
 with the standard deviation 
\begin{equation} 
\sigma_1 = \sqrt{\frac{\sum\limits_{i=1}^{N} (\Delta E_i - \overline {\Delta E})^2}{(N-1)}} \; , 
\end{equation}
is found to be $\overline {\Delta E}  \pm \sigma_1 = -566 \pm 548$ cm$^{-1}$. That corresponds to an average relative difference 
of $ \overline {\Delta x} \pm \sigma_2 = -0.009\% \pm 0.016\%$, with
\begin{equation}
\overline {\Delta x} \equiv \frac{\sum \limits_{i=1}^{N} \Delta x_i}{N} =\frac{\sum \limits_{i=1}^{N} (E^i_{\rm MBPT}/E^i_{\rm MCDHF/RCI} - 1)}{N}  \; ,
\end{equation}
and 
\begin{equation}
\sigma_2 = \sqrt{\frac{\sum\limits_{i=1}^{N} (\Delta x_i- \overline {\Delta x})^2}{(N-1)}} \; .
\end{equation}
The maximum difference is $-$2496 cm$^{-1}$ for state $\#183 / 2s 2p^4(^2\!P)3p~^{1}\!P_1$ corresponding to about $-0.03\%$. The previous theoretical energies of ~\citet{Landi.2006.V640.p1171}, ~\citet{Butler.2008.V489.p1369} and~\citet{Nahar.2011.V97.p403} depart from our MCDHF/RCI values with average differences of $-81 \pm7~681$  cm$^{-1}$, $16~380 \pm5~674$  cm$^{-1}$, and $75~509 \pm 29~681$ cm$^{-1}$, respectively. The largest deviations from our MCDHF/RCI values are, respectively, 25~898 cm$^{-1}$ ($\#183 / 2p^6~^{1}\!S_0$), 30~764 cm$^{-1}$ and 97~145 cm$^{-1}$ ($\#200 / 2s 2p^4(^2\!P)3d~^{1}\!D_2$).

Most of the compiled NIST and CHIANTI energies listed in Table~\ref{tab.lev.Fe} show a good agreement with the present two data sets, except for a few states where the deviations from our values are larger than 10~000 cm$^{-1}$. Furthermore, although the NIST ASD and CHIANTI database are claimed to be
critically evaluated,  discrepancies occur for some states. For these states our calculated values show good agreement with the results from the NIST database, thus resolving the inconsistencies. For example, the NIST compiled values for $\#12 / 2s^22p^3(^4\!S)3s~^{3}\!S_1$ and $\#25 / 2s^22p^3(^2\!P)3s~^{3}\!P_2$ agree well with the present calculations, while the corresponding values from the CHIANTI database deviate from our results by over 10~000 cm$^{-1}$. For these states the CHIANTI observations seem to be wrong or at least affected by large uncertainties. 
To provide some insight in the difficulties of getting reliable 
excitation energies we look at the  $2s^22p^3(^4\!S)3s~^{3}\!S_1$ 
state in more detail. 
With the aid of the known energy of the $2s^22p^4~^3\!P_2$ state, the CHIANTI energy 6~670~224 cm$^{-1}$ for  $\#12 / 2s^22p^3(^4\!S)3s~^{3}\!S_1$ is obtained using the wavelength 14.992 \AA~of the observation $2s^22p^3(^4\!S)3s~^{3}\!S_1-2s^22p^4~^3\!P_2$. This CHIANTI wavelength provided by~\citet{Landi.2005.V160.p286}, is about 0.15\% higher than the NIST, MCDHF/RCI, and MBPT values (14.966 \AA, 14.969 \AA, and 14.969 \AA), but is much closer to the NIST, MCDHF/RCI, and MBPT results (14.995 \AA, 14.997 \AA, and 14.997 \AA) for the $2s^22p^3(^2\!D)3s~^{1}\!D_2-2s^22p^4~^1\!D_2$ transition.~\citet{Landi.2005.V160.p286} also pointed out that the CHIANTI energy for $2s^22p^3(^4\!S)3s~^{3}\!S_1$ was derived from a blended line of  the $2s^22p^3(^4\!S)3s~^{3}\!S_1-2s^22p^4~^3\!P_2$ and $2s^22p^3(^2\!D)3s~^{1}\!D_2-2s^22p^4~^1\!D_2$ transitions. That line blending in the spectral observations compiled by the CHIANTI database is most likely responsible for the large deviation with the present theoretical values. 

Compiled values from the NIST ASD are also questionable for a few $n = 3$ states of \ion{Fe}{19}. The values 6~923~000~cm$^{-1}$ for $\#21 /2s^2 2p^3 (^2P)3s~^3\!P_1$, 
7~450~000 cm$^{-1}$ for $\#78 / 2s^2 2p^3 (^2P)3d~^3\!F_3$, 7~567~000 cm$^{-1}$ for 
$\#86 / 2s^2 2p^3 (^2P)3d~^3\!P_1$, 7~554~000~cm$^{-1}$ 
for $\#90 / 2s^2 2p^3 (^2P)3d~^3\!D_2$, and 7~606~000~cm$^{-1}$ for $\#92 / 2s^2 2p^3 (^2P)3d~^1\!P_1$ do not have obvious counterparts in the present MCDHF/RCI and MBPT calculations. All these level energies have been included in Table~\ref{tab.lev.nistmbptlardif}.

To further assess the accuracy of our calculated energies, a comparison between the present MCDHF/RCI and MBPT values is carried out along the sequence with $Z=24-30$. The compiled values from the NIST ASD are also included in the comparison. Good agreement between the present MCDHF/RCI and MBPT data is obtained, and the absolute average differences with the standard deviations decrease from $-623 \pm 606$ cm$^{-1}$ for \ion{Cr}{17} to $-397 \pm 536$ cm$^{-1}$ for \ion{Zn}{23}, corresponding to the average differences from $-0.012\% \pm 0.021\%$ to $-0.009\% \pm 0.016\%$. Looking more carefully at the differences between the present two energy sets, we can observe that they are large for the $2s 2p^4 (^2P)3p~^1\!P_1$ and $2s 2p^4 (^2P)3d~^1\!D_2$ states in the sequence, and that the maximum differences decrease from $-2872$ cm$^{-1}$ and $-2699$ cm$^{-1}$ for \ion{Cr}{17} to $-1580$ cm$^{-1}$ and $-1830$ cm$^{-1}$ for \ion{Zn}{23}. These states have specific radial  characteristics compared with all other states arising from the same $2s 2p^4 3p$ and $2s 2p^4 3d$ configurations. An EOL calculation, which weights all states by a statistical weight factor of $2J +1$, is strongly dominated by these other states and therefore produces
radial orbitals less suited to describe the $^1\!P_1$ and $^1\!D_2$ states.

\begin{figure*}
	\plotone{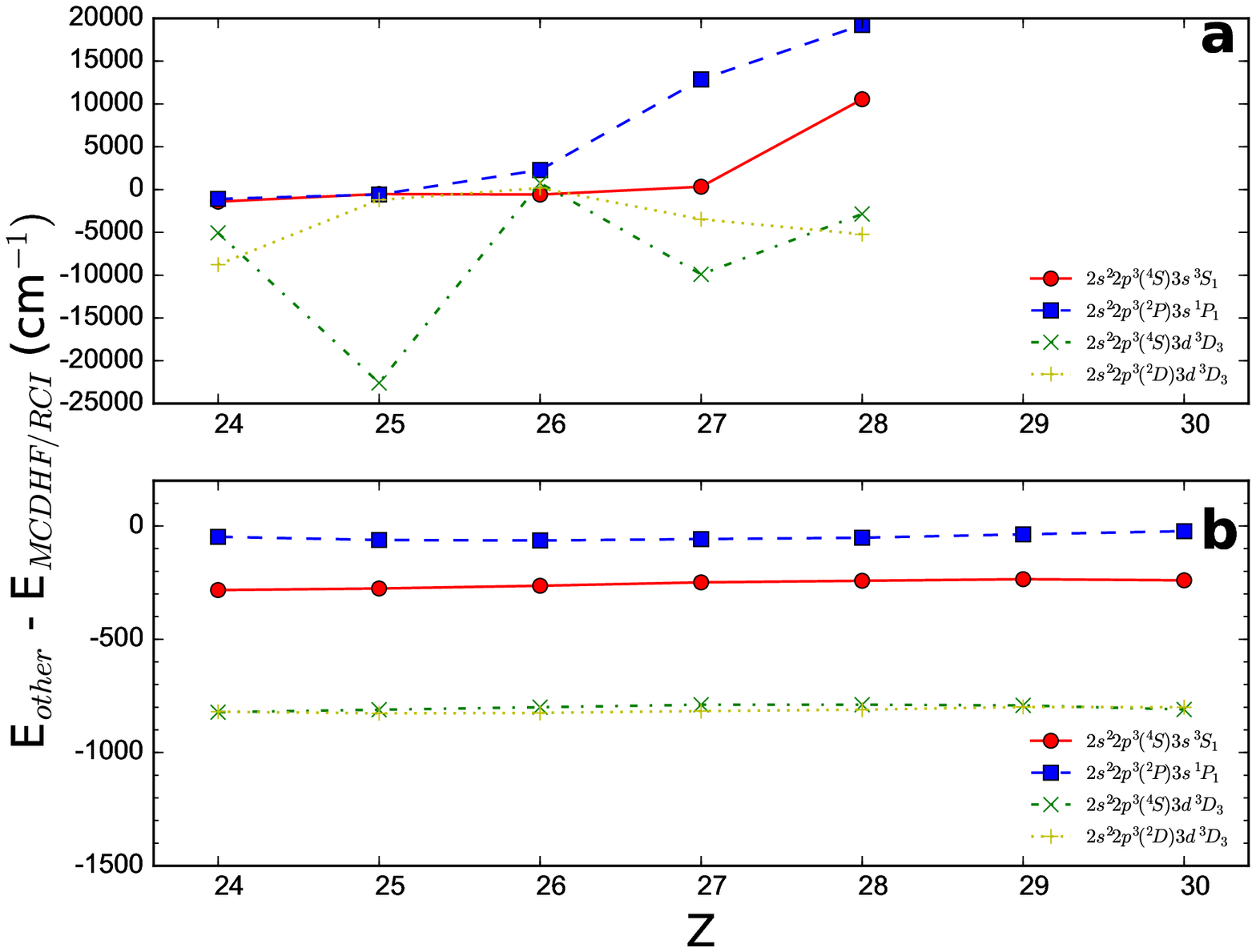}
	\caption{Energy deviations as a function of $Z$ for some levels:  (a) $E_{\rm NIST}-E_{\rm MCDHF/RCI}$ and (b) $E_{\rm MBPT}-E_{\rm MCDHF/RCI}$.\label{fig.lev.nistlargedifferences}}
\end{figure*}

For the states of the seven O-like ions considered in our work, the 149  compiled energies listed in the NIST ASD are all included in Table~\ref{tab.lev}. The NIST compiled values agree well with the present calculations for a majority of the states. However, there are about 41 states for which the NIST compiled values differ from our MCDHF/RCI results by over 4000 cm$^{-1}$ (0.05\%). All these 41  values and their reference sources have been listed in Table~\ref{tab.lev.nistmbptlardif}. These NIST compiled results should be reevaluated and used with care. As an example, in Figure~\ref{fig.lev.nistlargedifferences} (a) we show the deviations of the NIST energies to the present MCDHF/RCI results for some states as a function of $Z$. The deviations between the MBPT and MCDHF/RCI values for the same states along the sequence are shown in Figure~\ref{fig.lev.nistlargedifferences} (b). A few of the NIST compiled values depart from the MCDHF/RCI data by over than 10~000 cm$^{-1}$, while good agreement is obtained between the present two data sets. Moreover the differences of two data sets vary smoothly along the isoelectronic sequence. Therefore, these values compiled by the NIST ASD, which have been listed in Table~\ref{tab.lev.nistmbptlardif}, seem to be wrong or at least are affected by large errors.

\subsection{Radiative Rates}
In Table~\ref{tab.tr.sub}, transition wavelengths, transition rates $A$, weighted oscillator strengths $gf$, and line strengths $S$ are reported for O-like ions from \ion{Cr}{17} to \ion{Zn}{23}. Transition data in the length form for the present MCDHF/RCI and MBPT calculations are listed for the E1, M1, and E2 transitions connecting the present 200 levels with $A$ values larger than a fraction $10^{-3}$ of the sum of the $A$ values for the transitions from the upper level, i.e., radiative branching ratios (BRs) larger than 0.1~\%.

\begin{figure*}
	\plotone{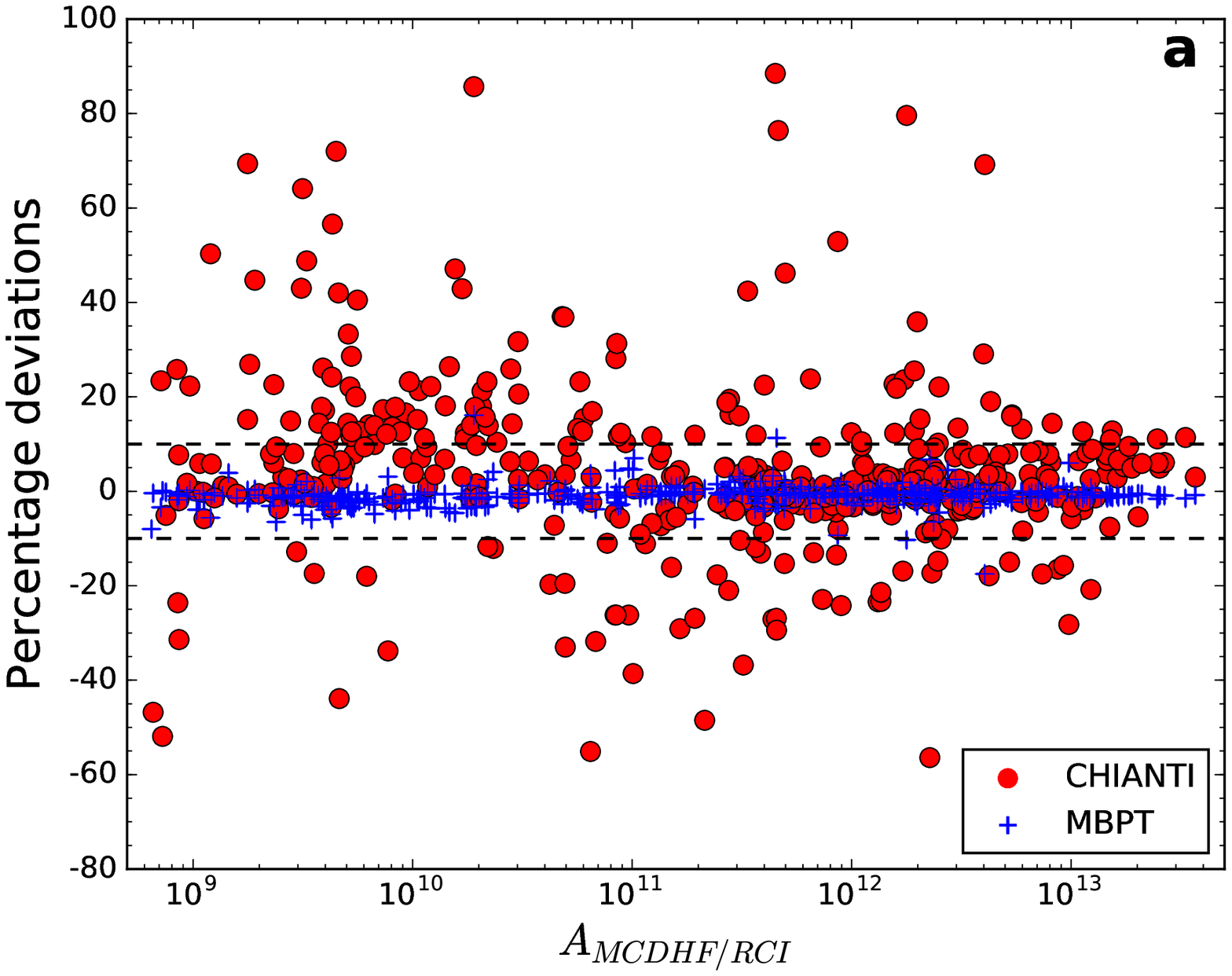}
	\plotone{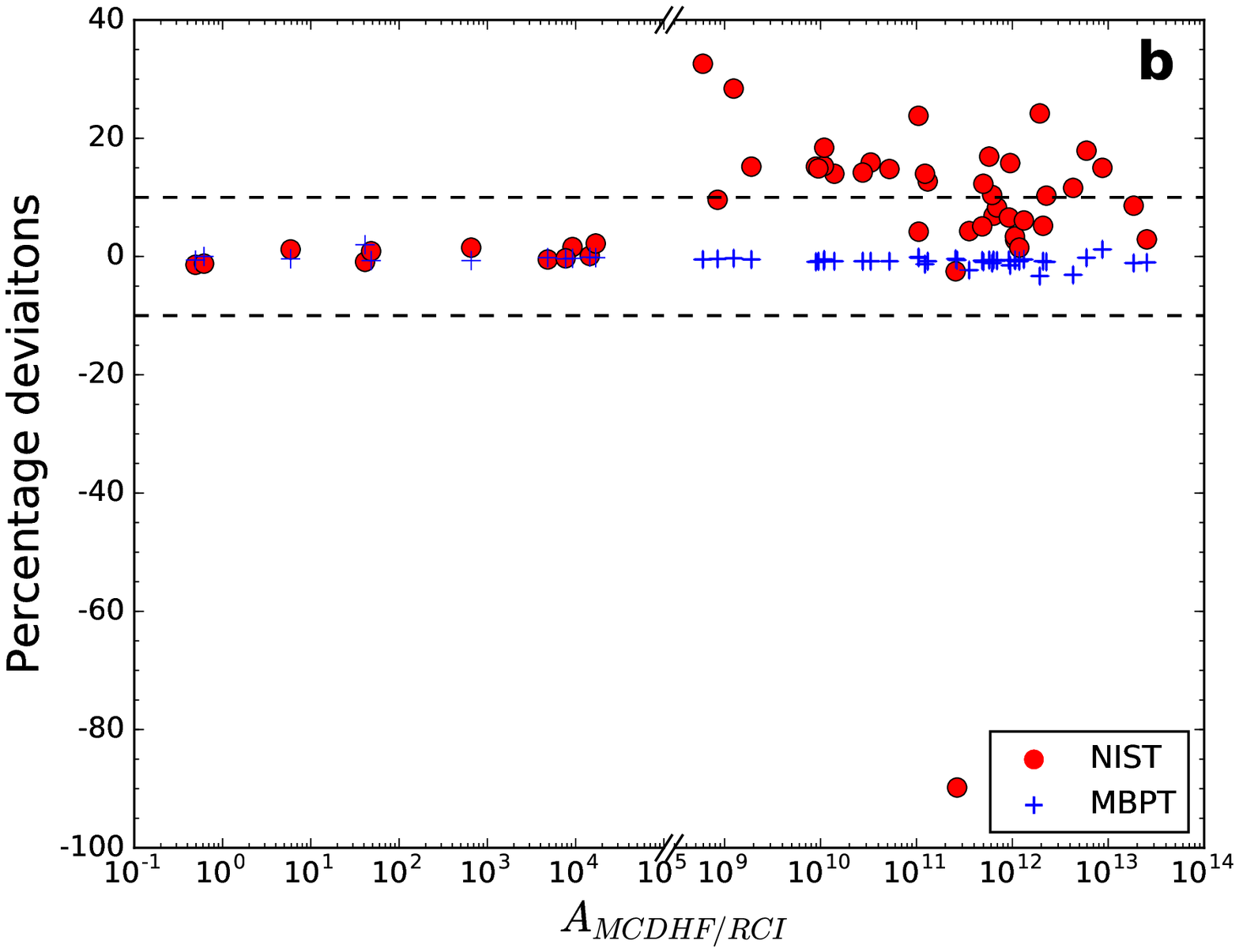}	
	\caption{(a) Percentage deviations from the CHIANTI and MBPT rates to the MCDHF/RCI rates for the transitions with BRs greater than 1~\% in \ion{Fe}{19}. (b) Percentage deviations from the NIST and MBPT rates to the MCDHF/RCI rates for the transitions in \ion{Fe}{19} listed by the NIST ASD. Dashed lines indicate the $\pm 10$~\% deviations. \label{fig.tr.chianti.nist}}
\end{figure*}

Since the transition data for iron are of particular interest in astrophysics, we compare the present MCDHF/RCI and MBPT rates for Fe XIX with rates from the CHIANTI and NIST databases. In Figure~\ref{fig.tr.chianti.nist}(a) the percentage deviations from the MBPT and CHIANTI $A$ values to the MCDHF/RCI rates for the transitions with BRs greater than 1~\% are shown. Many CHIANTI values differ from the MCDHF/RCI rates by 10\%-100\%, whereas the MBPT values agree with the MCDHF/RCI calculations to within 10\%. The average difference and standard deviation of the MBPT and MCDHF/RCI calculations is $-1~\% \pm 2~\%$. The corresponding  result from the CHIANTI values to the MCDHF/RCI rates is $2~\% \pm 15~\%$. The results of the comparisons can explained by the fact that the MBPT and MCDHF/RCI calculations consider more electron correlation effects than the CHIANTI values, which are reported by~\citet{Landi.2006.V640.p1171} using the standard RCI method. The CHIANTI values are considered to be less accurate compared with the present two data sets. 

In Figure~\ref{fig.tr.chianti.nist} (b), transition rates recommended by the NIST ASD and the corresponding MBPT rates are also compared with the MCDHF/RCI $A$ values. The 
rates of the present calculations agree within 1~\% for most of the transitions, whereas the NIST rates differ from the present calculations by 10~\%-90~\% for many transitions. For example, the MCDHF/RCI and MBPT rates for the $2s^2 2p^3 (^2D)3s~^1\!D_2 - 2s^2 2p^4~^3\!P_1$ transition are $2.66 \times 10^{11}$~s$^{-1}$ and $2.64 \times 10^{11}$ s$^{-1}$, respectively, while the corresponding NIST  value is about one order of magnitude smaller ($2.7 \times 10^{10}$ s$^{-1}$).

The above analysis leads us to the conclusion that compared with the CHIANTI and NIST transition data, 
our MCDHF/RCI and MBPT transition data are more accurate. As seen in Figure~\ref{fig.tr.chianti.nist}, the CHIANTI and NIST values differ from the MCDHF/RCI and MBPT rates relatively significantly, even for many strong transitions.  Using many $A$ values with less accuracy, particularly for strong transitions, to perform line identification or plasma modeling in astrophysics, quite different or even wrong results may be obtained. We highly recommend that the present MCDHF/RCI and MBPT transition data are used to update the CHIANTI and NIST data sets.

\begin{figure*}[h]
	\epsscale{1.05}
	\plotone{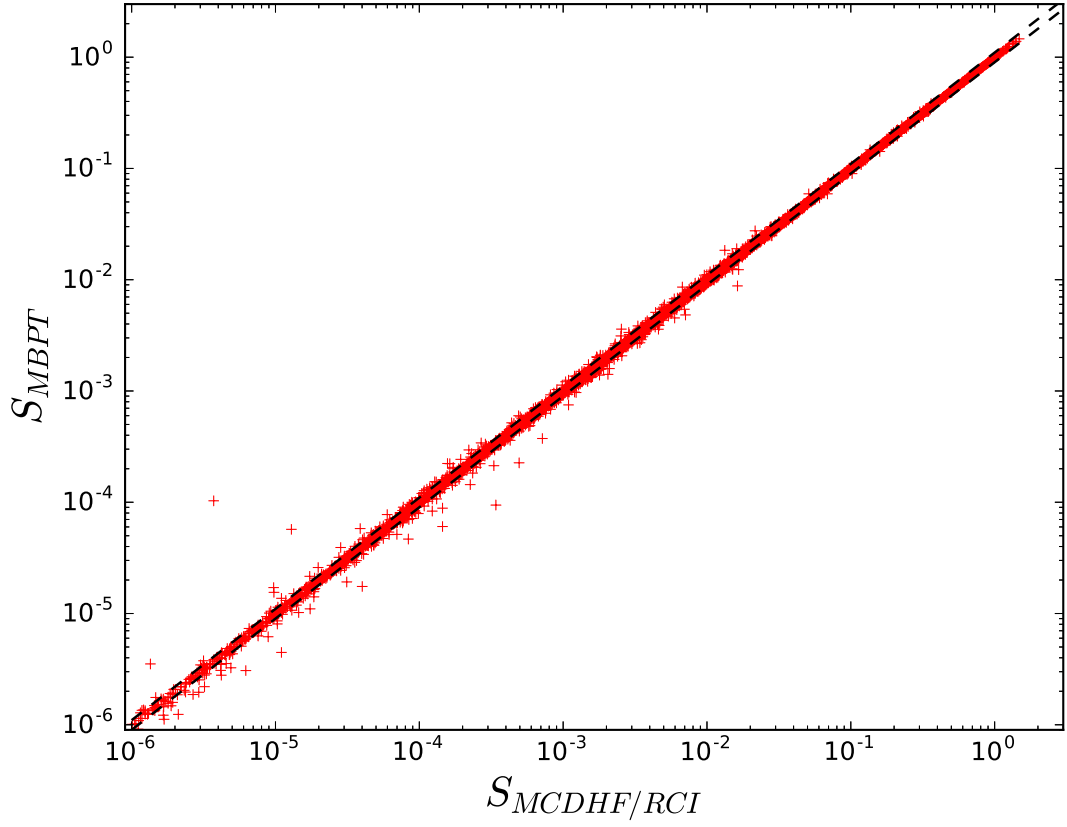}
	\caption{Comparison of the line strengths from our MCDHF/RCI ($S_{MCDF/RCI}$) and MBPT ($S_{MBPT}$) calculations for the E1 transitions. Dashed lines indicate the $\pm 10\%$  deviations. \label{fig.lns.mbpt.mcdf.e1}}
\end{figure*}

To further estimate the uncertainty of our transition data, the line strengths from our MCDHF/RCI calculations ($S_{MCDF/RCI}$) for the E1 transitions are compared with the present MBPT line strengths ($S_{MBPT}$) in Figure~\ref{fig.lns.mbpt.mcdf.e1}. Our two data sets agree within 10~\% for most of the transitions.
According to the uncertainty estimation method suggested by~\cite{Kramida.2014.V212.p11} we have the following averaged uncertainties for the $S$ values of E1 transitions in various ranges of the line strengths: 2~\% for $S \geq 10^{-1}$; 4~\% for $10^{-1} > S \geq 10^{-2}$; 6~\% for $10^{-2} > S \geq 10^{-3}$; 11~\% for $10^{-3} > S \geq 10^{-4}$; 22~\% for $10^{-4} > S \geq 10^{-5}$; and 30~\% for $10^{-5} > S \geq 10^{-6}$. Considering the contribution from the uncertainty of the wavelengths, about 11.1~\% transitions included in Table~\ref{tab.tr.sub} have $A$-value uncertainties of $\leq$ 3~\% (categories A$^{+}$ $\leq$ 2~\% and A $\leq$ 3~\% in the terminology of the NIST ASD), 61.3~\% have uncertainties of  $\leq$ 7~\% (category B$^{+}$),  2~\% have uncertainties of  $\leq$ 10~\% (category B), 15.5~\% have uncertainties of  $\leq$ 18~\% (category C$^{+}$), 7.3\% have uncertainties of  $\leq$ 25~\% (category C), 2.5~\% have uncertainties of  $\leq$ 40~\% (category D$^{+}$), and only 0.3~\% have uncertainties of  $>$ 40~\% (categories D and E). The uncertainty estimates of $A$ values for each transition are listed in the last column of Table~\ref{tab.tr.sub}.

Again, using the method suggested by~\cite{Kramida.2014.V212.p11}, the uncertainties of the $A$ values for the M1 and E2 transitions  have been estimated. They are listed in Table~\ref{tab.tr.sub} for each transition with BRs larger than 0.1~\%.

\subsection{Lifetimes}\label{sect:lt}
Our MCDHF/RCI and MBPT lifetimes are reported in Table~\ref{tab.lev}, including the contribution from all possible E1, M1, and E2 radiative rates from the corresponding states. 

Lifetimes for the first excited state $^3P_0$ of the configuration $2s^2 2p^4$ are dominated by the E2 transition $2s^2 2p^4~^3\!P_0~-~^3\!P_2$. Lifetimes for the other states $^3\!P_1$, $^1\!D_2$, and $^1\!S_0$ of the first excited configuration are determined by the M1 transitions to the ground state $^3\!P_2$ or to the state $^3\!P_1$. The E2 $A$ values from the states $2s^2 2p^3 3p, 3d$ to the states $2s^2 2p^4$ are important and contribute to the lifetimes for 10-40~\%. For the other states considered in the present calculations, lifetimes are mostly determined by the E1 transitions.  The present MCDHF/RCI and MBPT lifetimes agree to within 5\% for most states. The two sets of level lifetimes with large deviations mostly occur for the transitions with large cancellation effects. Even a slight difference in the calculations may lead to a relatively large deviations for transition rates in these cases. More detailed discussion on these effects can be found in our recent work~\citep{Si.2016.V227.p16}. 

\section{SUMMARY}
Employing two state-of-the-art methods (MCDHF/RCI and MBPT), the excitation energies and lifetimes of the lowest 200 states for the $2s^2 2p^4$, $2s 2p^5$, $2p^6$, $2s^2 2p^3 3s$, $2s^2 2p^3 3p$, $2s^2 2p^3 3d$, $2s 2p^4 3s$, $2s 2p^4 3p$, and $2s 2p^4 3d$ configurations have been calculated for O-like ions from \ion{Cr}{17} to \ion{Zn}{23}. Wavelengths, line strengths, transition rates, and oscillator strengths for
the E1, M1, and E2 transitions with BRs larger than 0.1~\% are also reported. 

There is a very good agreement between our MCDHF/RCI and MBPT results, and the absolute average energy differences with the standard deviations decrease from $-623 \pm 606$ cm$^{-1}$ for \ion{Cr}{17} to $-397 \pm 536$ cm$^{-1}$ for \ion{Zn}{23}, corresponding to the average differences from $-0.012\% \pm 0.021\%$ to $-0.009\% \pm 0.016\%$. Lifetimes agree to within 5\% for most states.
Observed values listed in Table~\ref{tab.lev.nistmbptlardif} compiled by the NIST ASD seem to be wrong or at least are affected by large errors.  The present calculations provide  a consistent and accurate data set for line identification and modeling purposes, which can also be considered as a benchmark for other calculations.

\acknowledgments
We acknowledge the support from the National Natural Science Foundation of China (Grant Grant No. 11674066, No.~21503066, No.~11504421, and No.~11474034) and the Project funded by the China Scholarship Council (Grant No. 201608130201). This work is also supported by the Chinese Association of Atomic and Molecular Data,  Chinese National Fusion Project for ITER No. 2015GB117000, the Swedish research council under contract 2015-04842 and the Belgian F.R.S.-FNRS Fonds de la Recherche Scientifique under CDR~J.0047.16. The authors (K.W. and S. L.) express their gratefully gratitude to the support from the visiting researcher program at the Fudan University.

\section*{Scientific software packages}
Scientific software packages including~\software{FAC \citep{Gu.2008.V86.p675},
	GRASP2K \citep{Jonsson.2007.V177.p597,Jonsson.2013.V184.p2197}} are used in the present work. We thanks the authors of these codes for providing support and guidance in using their codes.

\clearpage
\bibliographystyle{aasjournal}
\bibliography{ref}


\clearpage
\listofchanges

\clearpage

\clearpage

\clearpage
\begin{deluxetable*}{rrlrrrrrrl}
	\tabletypesize{\scriptsize}
\setlength{\tabcolsep}{4pt}
\tablecaption{Excitation energies (in cm$^{-1}$) and lifetimes ($\tau$ in s) for the lowest 200 levels of O-like ions with $Z=24-30$. Subscripts MCDHF-\\/RCI and MBPT represent the theoretical values from our MCDHFR/RCI and MBPT calculations, respectively; subscript NIST represents\\ the compiled values from the NIST ASD~\citep{Kramida.2015.V.p}. 
	The rightmost column gives the $LSJ$ coupling expansion coefficients of\\ the states indicated by the keys in parenthesis.\label{tab.lev}}
\tablehead{$Z$ & Key & Configuration & Term & $E_{\rm NIST}$ & $E_{\rm MCDHF/RCI}$ & $E_{\rm MBPT}$ & $\tau_{\rm{MCDHF/RCI}}$ & $\tau_{\rm{MBPT}}$ & Coupling Expansion Coefficients}
\startdata
  26&   1& $2s^{2}2p^{4}         $& $^3\!P_2$& 0.000000E+00&  0.000000E+00& 0.000000E+00&    \nodata&    \nodata&   0.95 (  1)  0.31(  4)                         \\
26&   2& $2s^{2}2p^{4}         $& $^3\!P_0$& 7.525000E+04&  7.532453E+04& 7.525161E+04&  2.032E+00&  2.012E+00&   0.89 (  2)  0.44(  5)                         \\
26&   3& $2s^{2}2p^{4}         $& $^3\!P_1$& 8.944100E+04&  8.943628E+04& 8.953510E+04&  6.887E-05&  6.882E-05&  -1.00 (  3)                                    \\
26&   4& $2s^{2}2p^{4}         $& $^1\!D_2$& 1.688520E+05&  1.689795E+05& 1.688682E+05&  5.696E-05&  5.683E-05&   0.95 (  4) -0.31(  1)                         \\
26&   5& $2s^{2}2p^{4}         $& $^1\!S_0$& 3.251400E+05&  3.254364E+05& 3.252158E+05&  7.176E-06&  7.148E-06&  -0.89 (  5)  0.45(  2)                         \\
26&   6& $2s    2p^{5}         $& $^3\!P_2$& 9.228900E+05&  9.229189E+05& 9.222972E+05&  2.258E-11&  2.242E-11&  -1.00 (  6)                                    \\
26&   7& $2s    2p^{5}         $& $^3\!P_1$& 9.847400E+05&  9.847962E+05& 9.842101E+05&  1.887E-11&  1.875E-11&   0.98 (  7)                                    \\
26&   8& $2s    2p^{5}         $& $^3\!P_0$& 1.030020E+06&  1.030088E+06& 1.029531E+06&  1.910E-11&  1.897E-11&   1.00 (  8)                                    \\
\enddata
	\tablecomments{ Table~\ref{tab.lev} is published in its entirety in the machine-readable format. The values for Fe XIX are shown here for guidance regarding its form and content.}
\end{deluxetable*}

\clearpage
\begin{deluxetable*}{rlrrrrrrrr}
	\tabletypesize{\footnotesize}
	\setlength{\tabcolsep}{4pt}
	\tablecaption{Energy levels (in cm$^{-1}$) relative to the ground state for the lowest 200 states of \ion{Fe}{19}. For brevity, energies other than the present MCDHF values are listed as differences from the latter ones in cm$^{-1}$. $E_{\rm MCDHF/RCI}$, $E_{\rm MBPT}$-the present calculations; $E_{\rm NIST}$-\citet{Kramida.2015.V.p}; $E_{\rm{CHIANTI}}$-\citet{Dere.1997.V125.p149,DelZanna.2015.V582.p56}; $E_{\rm{Landi}}$-\citet{Landi.2006.V640.p1171}; $E_{\rm{Butler}}$-\citet{Butler.2008.V489.p1369}; $E_{\rm{Nahar}}$-\citet{Nahar.2011.V97.p403}.\label{tab.lev.Fe}}
	\tablehead{\colhead{Key} & \colhead{Configuration} & \colhead{Term} & \colhead{$E_{\rm MCDHF}$} & \colhead{$E_{\rm MBPT}$} & \colhead{$E_{\rm NIST}$} & \colhead{$E_{\rm{CHIANTI}}$} & \colhead{$E_{\rm{Landi}}$} & \colhead{$E_{\rm{Butler}}$} &  \colhead{$E_{\rm{Nahar}}$}}
	\startdata
$1  $& $2s^{2}2p^{4}         $& $^3\!P_2$&       $      0   $&$0    $&$0      $&$0         $&$0         $&$0    $&$0     $\\
$2  $& $2s^{2}2p^{4}         $& $^3\!P_0$&       $  75325   $&$-73  $&$-75    $&$-75       $&$-126      $&$-946 $&$-45   $\\
$3  $& $2s^{2}2p^{4}         $& $^3\!P_1$&       $  89436   $&$99   $&$5      $&$5         $&$-616      $&$29   $&$0     $\\
$4  $& $2s^{2}2p^{4}         $& $^1\!D_2$&       $ 168980   $&$-111 $&$-128   $&$-128      $&$1778      $&$4810 $&$-94   $\\
$5  $& $2s^{2}2p^{4}         $& $^1\!S_0$&       $ 325436   $&$-221 $&$-296   $&$-296      $&$-16       $&$3186 $&$-285  $\\
$6  $& $2s    2p^{5}         $& $^3\!P_2$&       $ 922919   $&$-622 $&$-29    $&$-29       $&$6312      $&$8283 $&$-28   $\\
$7  $& $2s    2p^{5}         $& $^3\!P_1$&       $ 984796   $&$-586 $&$-56    $&$-56       $&$6450      $&$9451 $&$-57   $\\
$8  $& $2s    2p^{5}         $& $^3\!P_0$&       $1030088   $&$-557 $&$-68    $&$-68       $&$5970      $&$9665 $&$-72   $\\
\enddata
\end{deluxetable*}
\clearpage

\begin{deluxetable*}{lclccccccl}
	\tabletypesize{\tiny}
\setlength{\tabcolsep}{4pt}
	\tablecaption{{\textnormal {Excitation energies (in cm$^{-1}$) for the states for which the NIST compiled values differ from our MCDHF/RCI values by more than 4000 cm$^{-1}$ (0.05\%), as well as their reference sources.}} \label{tab.lev.nistmbptlardif}}
	\tablehead{\colhead{\emph{Z}} &\colhead{Key\tablenotemark{a}}  &\colhead{Configuration} & \colhead{Term}  & \multicolumn{3}{c}{ Energy } &\multicolumn{2}{c}{ $\Delta E$} &\multicolumn{1}{c}{ References}\\
		\cmidrule(r){5-7}	\cmidrule(r){8-9} 
		& & &&\colhead{MCDHF/RCI\tablenotemark{b}}&\colhead{MBPT\tablenotemark{c}} &\colhead{NIST\tablenotemark{d}} &\colhead{MBPT\&MCDHF/RCI} &\colhead{NIST\&MCDHF/RCI}\vspace{ -4mm} \\}
	\startdata
24	&54	& $2s^2 2p^3(^4\!S)3d$&$^3\!D_3$&  5953561&  5952739& 5948500&-822 &  -5061  &~\citep{Shirai.2000.V.p}\\
24	&68	& $2s^2 2p^3(^2\!D)3d$&$^3\!D_3$&  6082766&  6081946& 6074000&-820 &  -8766  &~\citep{Shirai.2000.V.p}\\
24	&75	& $2s^2 2p^3(^2\!D)3d$&$^1\!F_3$&  6129342&  6128282& 6124400&-1060&  -4942  &~\citep{Shirai.2000.V.p}\\
24	&85	& $2s^2 2p^3(^2\!P)3d$&$^3\!P_2$&  6181128&  6180748& 6131000&-380 &  -50128 &~\citep{Fawcett.1975.V170.p185,Shirai.2000.V.p}\\
24	&87	& $2s^2 2p^3(^2\!P)3d$&$^3\!D_3$&  6199220&  6198792& 6164800&-428 &  -34420 &~\citep{Fawcett.1975.V170.p185,Shirai.2000.V.p}\\
24	&90	& $2s^2 2p^3(^2\!P)3d$&$^3\!D_2$&  6218770&  6218042& 6214600&-728 &  -4170  &~\citep{Fawcett.1975.V170.p185,Shirai.2000.V.p}\\
25	&54	& $2s^2 2p^3(^4\!S)3d$&$^3\!D_3$&  6585083&  6584272& 6562500&-811 &  -22583 &~\citep{Fawcett.1975.V170.p185,Shirai.2000.V.p}\\
25	&85	& $2s^2 2p^3(^2\!P)3d$&$^3\!P_2$&  6835162&  6834780& 6778700&-382 &  -56462 &~\citep{Fawcett.1975.V170.p185,Shirai.2000.V.p}\\
\enddata
	\tablenotetext{a}{The index number of the level given in Table~\ref{tab.lev}.}
	\tablenotetext{b}{Our MCDHF/RCI energies.}	
	\tablenotetext{c}{Our MBPT energies.}	
	\tablenotetext{d}{The NIST recommended energies \citep{Kramida.2015.V.p}.}
\end{deluxetable*}
\clearpage
\clearpage
\clearpage
\begin{deluxetable*}{rrrcccccccccl}
	\tabletypesize{\scriptsize}
	\tablecolumns{7}
	\tablewidth{0pt}
	\tablecaption{{\textnormal {Transition data for O-like ions from \ion{Cr}{17} to \ion{Zn}{23}. $\lambda$ is transition wavelength in {\AA}, \emph{S} is line strength in atomic units, $gf$ is weighted oscillator strength, and \emph{A} is transition rate in s$^{-1}$. Only transitions which contribute to the total radiative lifetime by more than 0.1\% are presented. The last column represents the estimated accuracies of the $A$-values using the terminologies of the NIST ASD~\citep{Kramida.2015.V.p}.}} \label{tab.tr.sub}}
	\tablehead{$Z$ & \emph{j} & \emph{i} & Type & $\lambda_{\rm MCDHF}$ & $\lambda_{\rm MBPT}$ & $\emph{S}_{\rm MCDHF}$ & \emph{gf}$_{\rm MCDHF}$  & $A_{\rm MCDHF}$  & $S_{\rm MBPT}$ & \emph{gf}$_{\rm MBPT}$ & $A_{\rm MBPT}$ & Acc. \vspace{ -4mm} \\}
	\startdata
24&  2 &   1&   E2& 1.7207E+03&  1.7194E+03&  2.505E-03&  8.276E-11&  1.867E-01&  2.486E-03&  8.193E-11&  1.846E-01&  B+\\ 
24&  3 &   1&   M1& 1.6541E+03&  1.6562E+03&  2.319E+00&  5.662E-06&  4.589E+03&  2.308E+00&  5.642E-06&  4.585E+03&  A \\ 
24&  4 &   1&   M1& 7.4090E+02&  7.4002E+02&  4.890E-01&  2.672E-06&  6.509E+03&  4.894E-01&  2.671E-06&  6.492E+03&  A \\ 
24&  4 &   3&   M1& 1.3420E+03&  1.3378E+03&  1.777E-01&  5.371E-07&  4.004E+02&  1.779E-01&  5.362E-07&  3.972E+02&  A \\ 
24&  5 &   3&   M1& 4.9370E+02&  4.9283E+02&  2.602E-01&  2.135E-06&  5.864E+04&  2.604E-01&  2.133E-06&  5.836E+04&  A \\ 
24&  6 &   1&   E1& 1.2307E+02&  1.2295E+02&  1.297E-01&  3.205E-01&  2.828E+10&  1.289E-01&  3.182E-01&  2.803E+10&  A+\\ 
24&  6 &   3&   E1& 1.3296E+02&  1.3281E+02&  4.601E-02&  1.052E-01&  7.959E+09&  4.576E-02&  1.045E-01&  7.888E+09&  B+\\ 
24&  6 &   4&   E1& 1.4758E+02&  1.4745E+02&  9.590E-03&  1.976E-02&  1.212E+09&  9.586E-03&  1.973E-02&  1.208E+09&  B+\\
\enddata
	\tablenotetext{\space}{Table~\ref{tab.tr.sub} is published in its entirety in the machine-readable format. A portion is shown here for guidance regarding its form and content.}
\end{deluxetable*}
\clearpage

\end{document}